\documentclass[12pt,showpacs,preprintnumbers,superscriptaddress,amsmath,amssymb,nofootinbib]{revtex4}
\usepackage{graphicx}% Include figure files
\usepackage{dcolumn}% Align table columns on decimal point
\usepackage{bm}% bold math
\usepackage{amssymb}
\usepackage{amsmath}
\usepackage{epsfig}    
\usepackage{color}
\usepackage{slashed}
\usepackage{hhline}
\def\be{\begin{equation}}
\def\ee{\end{equation}}
\newcommand{\bea}{\begin{eqnarray}}
\newcommand{\eea}{\end{eqnarray}}
\newcommand{\nn}{\nonumber}

\numberwithin{equation}{section}

\begin{document}

%%%%%%%%%
\title{3.55 keV X-ray Line Signal from Excited Dark Matter in
Radiative Neutrino Model 
}
\preprint{KIAS-P14027}
\preprint{IPPP-14-38}
\preprint{DCPT-14-76}

\author{Hiroshi Okada}
\email{hokada@kias.re.kr}
\affiliation{School of Physics, KIAS, Seoul 130-722, Korea}
\author{Takashi Toma}
\email{takashi.toma@durham.ac.uk}
\affiliation{Institute for Particle Physics Phenomenology\\
 University of
Durham, Durham DH1 3LE, United Kingdom} 

\begin{abstract}
We study an exciting dark matter scenario in a radiative neutrino model to explain
 the X-ray line signal at $3.55$ keV  
recently reported by XMN-Newton X-ray observatory using data of various
 galaxy clusters and Andromeda galaxy. 
We show that the required large cross section for the up-scattering
 process to explain the X-ray line can be obtained via the resonance of the pseudo-scalar. 
Moreover this model can be compatible with the thermal production of dark
 matter and the constraint from the direct detection experiment.  
\end{abstract}
\maketitle
\newpage

\section{Introduction}
%%%
In the light of anomalous X-ray line signal at  $3.55$ keV from the analysis of
 XMN-Newton X-ray observatory  data of various galaxy clusters and
 Andromeda galaxy~\cite{Bulbul:2014sua, Boyarsky:2014jta}, dark matter (DM)
 whose mass is in the range from keV to GeV comes into one of the
 promising candidates. Subsequently, a number of literatures are
 recently arising around the subject~\cite{Ky:2005yq, Dinh:2006ia,
 Merle:2013gea, Ishida:2014dlp, Finkbeiner:2014sja, 
Higaki:2014zua, Jaeckel:2014qea,Lee:2014xua, Kong:2014gea,
 Frandsen:2014lfa, Baek:2014qwa, Cline:2014eaa, Modak:2014vva,
 Babu:2014pxa, Queiroz:2014yna, Demidov:2014hka, Ko:2014xda,
 Allahverdi:2014dqa,  Kolda:2014ppa, Cicoli:2014bfa, Dudas:2014ixa, Choi:2014tva}. 
 As for the keV scale DM, for example, a sterile neutrino can be one of
 the typical candidates to explain the X-ray anomaly that requires tiny
 mixing between the DM and the active neutrino;
 $\sin^22\theta\approx10^{-10}$~\cite{Bulbul:2014sua}. However these scenarios suggest
 that neutrino masses cannot be derived consistently with the sterile
 neutrino DM due to its too small mixing. 
 Moreover, the sterile neutrino DM mass is out
 of the range in the direct detection searches such as
 LUX~\cite{Akerib:2013tjd}, which is currently the most powerful
 experiment to constrain the kind of Weakly Interacting Massive Particle. 

As for the GeV scale DM, on the other hand, the exciting DM
scenario which requests a pair of ground state and excited DM is
known to explain the X-ray~\cite{Finkbeiner:2014sja}. In this framework,
the emission of X-ray is simply realized as follows. 
After the ground state DM up-annihilates into the excited DM
pair, it can decay into photons (X-ray) and the ground state DM. The mass
difference among them is assumed to be the energy of the X-ray, 3.55
keV. Since the framework of the exciting DM is simple, 
this scenario can be applicable to various models such as radiative
neutrino models~\cite{Radseesaw-1-1,Radseesaw-1-2, Radseesaw-2,
Radseesaw-3}. 
In this kind of models, small neutrino masses and existence of DM would be accommodated 
unlike the sterile neutrino DM scenarios above. 
Moreover the DM can be testable in direct detection searches because the
DM mass is GeV scale. 
%%%

In this letter, we account for the X-ray anomaly in terms of an excited
DM scenario in a simple extended model with radiative neutrino
masses~\cite{Radseesaw-1-2}, in which 
three right-handed neutrinos, a $SU(2)_L$ doublet scalar and a singlet
scalar are added to the Standard Model (SM) and the first two lightest
right-handed neutrinos are assumed to be a pair of ground state and
excited state DM.

%%%%%%%%%%%%%%%%%%%%%%%%%%%%%%%%%%%%%%
\section{The Model}
\subsection{Model setup}

\begin{table}[thbp]
\centering {\fontsize{10}{12}
\begin{tabular}{|c|c|c||c|c|c|c|}
\hline Particle   & $L_i$   & $e_i$  & $N_i$   & $\eta$  & $\Phi$   & $\Sigma$
  \\\hhline{|=|=|=#=|=|=|=|}
$(SU(2)_L,U(1)_Y)$  & $(\bm{2},-1/2)$ & $(\bm{1},-1)$ & $(\bm{1},0)$ &
		 $(\bm{2},1/2)$  & $(\bm{2},1/2)$ & $(\bm{1},0)$ 
\\\hline
$Z_3$  & $\omega^2$ & $1$ & $\omega$  & $\omega^2$ & $\omega^2$ & $\omega$  \\\hline
%%%
$Z_2$  & $+$ & $+$ & $-$  & $-$ & $+$ & $+$  \\\hline
%%%
\end{tabular}%
} \caption{The new particle contents and the charges for bosons where $i=1-3$
 is generation index.}
\label{tab:1}
\end{table}
%%%
The particle contents and charge assignments of the model we consider
are shown in Tab.~\ref{tab:1}.
We introduce three right-handed neutrinos $N_i$ ($i=1-3$) where the
first two lightest ones are identified to be a pair of ground
state and excited state DM. We also introduce a $SU(2)_L$ doublet
inert scalar $\eta$ that is assumed not to have vacuum expectation value (VEV),
and a gauge singlet boson $\Sigma$ with non-zero VEV in addition to the
SM like Higgs boson $\Phi$. 
The $Z_2$ symmetry is imposed to assure the stability of DM. The $Z_3$
symmetry plays an important role 
in forbidding the term $(\Sigma+\Sigma^\dag)\overline{N^c_i}P_RN_i$ that leads
no pseudo scalar coupling like $\Sigma_I\overline{N_i^c}\gamma_5N_i$
where $\Sigma_I$ is the imaginary part of $\Sigma$. As
we will see later, the pseudo scalar 
coupling is important to induce up-scattering process $N_1N_1\to
N_2N_2$. 
The $Z_3$ symmetry also allows the cubic term $\Sigma^3+\mathrm{h.c.}$
that provides the mass of the pseudo scalar component of $\Sigma$.
%%%
The relevant Lagrangian for the discussion is given as follows
\begin{eqnarray}
\mathcal{L}
&=&
(D^\mu\Phi)^\dag(D_\mu\Phi)+(D^\mu\eta)^\dag(D_\mu\eta)
+\left(y_\ell \bar L \Phi e + y_\eta \bar L \eta^\dag N \nn
+\frac{y_N}{2} \Sigma \bar N^c N
+{\rm h.c.}\right), \label{KT}
\\
\mathcal{V}
&=& 
 m_1^{2} \Phi^\dagger \Phi + m_2^{2} \eta^\dagger \eta  + m_3^{2}
 \Sigma^\dagger \Sigma + \left(\mu\Sigma^3+\mathrm{h.c.}\right)
 \nn
\\ &&
+\lambda_1 (\Phi^\dagger \Phi)^{2} + \lambda_2 
(\eta^\dagger \eta)^{2} + \lambda_3 (\Phi^\dagger \Phi)(\eta^\dagger \eta)
+ \lambda_4 (\Phi^\dagger \eta)(\eta^\dagger \Phi)
+\left[\lambda_5(\Phi^\dagger \eta)^{2} + \mathrm{h.c.}\right]\nn\\
%%%%%%%
&&+\lambda_6 (\Sigma^\dagger \Sigma)^{2} + \lambda_7  (\Sigma^\dagger
\Sigma)(\Phi^\dagger \Phi) 
+ \lambda_8  (\Sigma^\dagger \Sigma) (\eta^\dagger \eta),
\label{HP}
\end{eqnarray}
where the generation indices are omitted, and 
the Yukawa coupling $y_N$ can be regarded as diagonal in general. 
After the electroweak symmetry breaking, the scalar fields can be parametrized as 
\begin{align}
&\Phi =\left(
\begin{array}{c}
G^+\\
\frac{v+\phi^0+iG^0}{\sqrt{2}}
\end{array}\right),
\quad
\eta =\left(
\begin{array}{c}
\eta^+\\
\frac1{\sqrt2}(\eta_R+i\eta_I)
\end{array}\right),\quad
\Sigma=\frac{v'+\sigma+i\rho}{\sqrt{2}},
\label{component}
\end{align}
where $ v\approx246$ GeV, and $G^+$ and $G^0$ are absorbed in $W^+$
boson and $Z$ boson due to the 
Higgs mechanism. The resulting CP even mass matrix with nonzero VEV is given by 
\begin{equation}
m^{2} (\phi^{0},\sigma) = \left(
\begin{array}{cc}
  2\lambda_1v^2 & \lambda_7vv'
  \\
  \lambda_7vv'
  &\frac{(3\sqrt{2}\mu+4\lambda_{6}v')v'}{2}\\
\end{array}
\right),
\end{equation}
where the tadpole conditions
$\left.\partial{\mathcal{V}}/\partial\phi^0\right|_{\mathrm{VEV}}=0$ and 
$\left.\partial{\mathcal{V}}/\partial\sigma\right|_{\mathrm{VEV}}=0$ are inserted.  
This mass matrix is diagonalized by the rotation matrix, and $\phi^0$ and
$\sigma$ are rewritten by the mass eigenstates $h$ and $H$ as
\begin{eqnarray}
\phi^0 &=& h\cos\alpha + H\sin\alpha, \nn\\
\sigma &=&- h\sin\alpha + H\cos\alpha.
\label{eq:mass_weak}
\end{eqnarray}
The mass eigenstate $h$ corresponds to the SM-like Higgs and $H$ is an
extra Higgs respectively. The mixing angle $\sin\alpha$ is expressed as the
function in terms of the other parameters as
\begin{eqnarray}
&&\sin 2\alpha=\frac{ \lambda_7vv' }{m_h^2-m_H^2}.
\end{eqnarray}
The pseudo scalar $\rho$ does not mix after the symmetry breaking and
the mass is just given by $m_\rho^2=9\mu^2/\sqrt{2}$.
The masses of the other $Z_2$ odd scalars $\eta^+$, $\eta_R$ and
$\eta_I$ are also determined adequately to be 
\begin{eqnarray}
m_\eta^2 &=& m_2^{2} + \frac12 \lambda_3 v^{2}
 + \frac12 \lambda_8 v'^{2}
 , \\ 
m^2_{{R}} &=& m_2^{2} + \frac12 \lambda_8 v'^{2}
 + \frac12 (\lambda_3 + \lambda_4 + 2\lambda_5) v^{2}
 , \\ 
m^2_{{I}} &=& m_2^{2} + \frac12 \lambda_8 v'^{2}
 + \frac12 (\lambda_3 + \lambda_4 - 2\lambda_5) v^{2}.
\label{eq:eta-mass}
\end{eqnarray}
The mass splitting between $m_R$  and $m_I$ is given by
$m_R^2-m_I^2=2\lambda_5 v^2$.
The lower bounds of the inert scalar masses are obtained as
$m_\eta\gtrsim70~\mathrm{GeV}$ and $m_R$, $m_I\geq45~\mathrm{GeV}$ by the LEP
experiment~\cite{Swiezewska:2012eh, Lundstrom:2008ai, ALEPH:2002aa} and
the invisible decay of $Z$ boson~\cite{ALEPH:2002aa}. 
In addition, the mass difference between the charged and neutral inert
scalars is constrained as roughly less than
$\mathcal{O}(100)~\mathrm{GeV}$ by the $T$ parameter~\cite{Barbieri:2006dq}.

\subsection{Neutrino sector}
The right-handed neutrinos obtain the masses after the
symmetry breaking due to VEV of $\Sigma$, 
\begin{equation}
M=\frac{v'}{\sqrt{2}}\left(
\begin{array}{ccc}
y_{N1} & 0      & 0\\
0      & y_{N2} & 0\\
0      & 0      & y_{N3}
\end{array}
\right)\equiv\left(
\begin{array}{ccc}
M_1 & 0   & 0\\
0   & M_2 & 0\\
0   & 0   & M_3
\end{array}
\right).
\end{equation}
Using the right-handed neutrino masses, the active neutrino masses can be
obtained at one-loop level as~\cite{Radseesaw-1-2}
\be
(m_\nu)_{ab}=\sum_i\frac{(y_\eta)_{ai}(y_\eta)_{bi}M_{i}}{2(4\pi)^2}
\left[\frac{m^2_R}{m^2_R-M^2_{i}}\ln\frac{m^2_R}{M^2_{i}}
-
\frac{m^2_I}{m^2_I-M^2_{i}}\ln\frac{m^2_I}{M^2_{i}}\right].
\ee
In particular, when the mass splitting between $\eta_R$ and $\eta_I$ is
small ($\lambda_5\ll1$) and $N_i$ are much lighter than $\eta$ ($M_i\ll
m_{R}\approx m_{I}$), the formula can be simplified as follows 
\be
(m_\nu)_{ab}\approx 
\frac{\lambda_5 v^2}{(4\pi)^2(m^2_R+m^2_I)}
\sum_i(y_\eta)_{ai}(y_\eta)_{bi}M_{i}.
\ee
We will consider the mass hierarchy for the analysis in the next section. 
The following parameter set is taken for example to be consistent with the sum of
the light neutrino masses 0.933 eV~\cite{Ade:2013lta} 
\be
M\sim\mathcal{O}(10)\ {\rm GeV},\quad y_\eta \approx 0.1,\quad\lambda_5\approx10^{-5},\quad
m_R\approx m_I\sim \mathcal{O}(1)\ {\rm TeV}.
\label{eq:ex-para}
\ee
Note that the Yukawa coupling $y_\eta$ cannot be too small since the lifetime
of the decay channel $N_2\to N_1\gamma$ becomes too long to explain the
X-ray anomaly. 

Lepton Flavor Violating processes such as $\mu\to
e\gamma$ or $\mu\to3e$ should be taken into account~\cite{Toma:2013zsa}. 
One may think that the above parametrization has been
already excluded by the strong constraint of $\mu\to e\gamma$ whose
branting ratio should be $\mathrm{Br}(\mu\to e\gamma)\leq5.7\times10^{-13}$. 
However it can be evaded by considering a specific flavor structure of
the Yukawa coupling $y_\eta$ as ref.~\cite{Suematsu:2009ww, Suematsu:2010gv, Schmidt:2012yg}.

%%%%%%%%%%%%%%%%%%%%%
\section{Dark Matter}
We identify that $N_1$ and $N_2$ are a pair of ground and excited state
DM for explaining the X-ray anomaly. Thus their masses are related as
$M_{1}\approx \ M_{2}< M_{3}$, and 
$M_{2}-M_{1}\equiv \Delta M=3.55~\mathrm{keV}$.
Such the situation has been considered for a different motivation in
ref.~\cite{Suematsu:2010gv, Suematsu:2009ww, Schmidt:2012yg}. 
The small mass splitting between $N_1$ and $N_2$ would be theoretically
derived by introducing an extra $U(1)$ symmetry. For example, we can
construct the model that the interactions $\Sigma N_1N_1$ and $\Sigma
N_2 N_2$ are forbidden but $\Sigma N_1N_2$ is allowed, and the small
$U(1)$ breaking terms such as $N_1N_1$ and $N_2N_2$ come from higher
dimensional operators. Then after diagonalizing the mass matrix composed
by $N_1$ and $N_2$, almost degenerated two mass eigenstates are obtained. 

A small momentum of DM is required to lead the up-scattering event
$N_1N_1\to N_2N_2$. To induce the up-scattering process, the
required minimum relative velocity of a pair of DM is 
estimated as $v_{\mathrm{min}}\approx2\sqrt{2\Delta M/M_{1}}$ from the
kinematics. 
It suggests that the mass of DM should be ${\cal O}$(10) GeV since the
averaged DM velocity in the present universe is estimated as $v_{\rm
rel}\sim 10^{-3}$.
The local photon flux is approximately  given by~\cite{Finkbeiner:2014sja} 
\begin{eqnarray}
F\approx 2.6\times10^{-5}
\left[\frac{\langle\sigma v_{\rm rel}(N_1N_1 \to N_2N_2)\rangle
 }{10^{-19}{\rm cm^3/s}}\right]
\left[\frac{10\ {\rm GeV}}{M_{1}}\right]^2\ {\rm photon/sec/cm^2},
\end{eqnarray}
where NFW profile is assumed~\cite{Navarro:1996gj}.
The above formula suggests that we need a quite large cross section to
explain the X-ray line. 
However our model can obtain such a large cross section via the $\rho$
resonance. 

\begin{figure}[tbc]
\begin{center}
\includegraphics[scale=0.8]{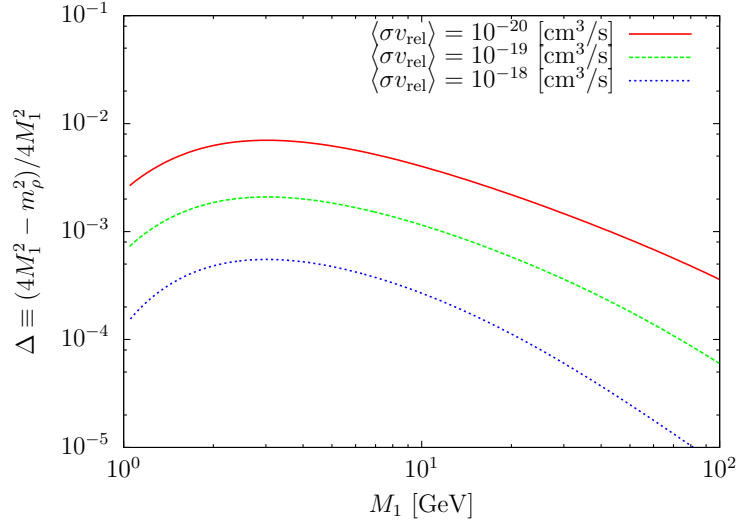}
   \caption{Contours of velocity averaged cross section in $M_1-\Delta$ plane. We find that
 rather mild fine tuning provides the required cross section over the
 range of DM mass depicted in figure.  Here we fix as $y_{N1}=y_{N2}\approx1$.}
   \label{mix-plot}
\end{center}
\end{figure}

The up-scattering process $N_1N_1\to N_2N_2$ is 
derived by the massive pseudo-scalar $\rho$. 
Since the required cross section for the process is very large as
$\sigma{v}_{\mathrm{rel}}\sim10^{-19}~\mathrm{cm^3/s}$~\cite{Finkbeiner:2014sja},
an enhancement 
mechanism is required\footnote{Note here that the pair of $\eta_I$ and
$\eta_R$ cannot be used to explain the X-ray line  
because the decay process $\eta_R\to\eta_I\gamma$ is forbidden by spin
statistics.}. 
In our case, we have the resonance in the $\rho$ mediated s-channel
which leads s-wave for the cross section. 
In this sense, the interaction between the pseudo scalar $\rho$ and two
DM are crucially important. 
The up-scattering cross section is given by
\begin{eqnarray}
&&\sigma v_{\rm rel}(N_1N_1\to N_2N_2)\approx
\frac{|y_{N1}y_{N2}|^2s}{16\pi }
\sqrt{1-\frac{4M^2_{2}}{s}} \frac{1}{(s-m^2_\rho)^2+\Gamma^2_\rho m^2_\rho} ,\\
&&\mbox{with}\quad
\Gamma_\rho=\frac{|y_{N1}|^2}{16\pi}m_\rho
\sqrt{1-\frac{4M^2_{1}}{m^2_\rho}},
\end{eqnarray}
where $s\approx 4M^2_{1}(1+v^2_{\rm rel}/4)$. 
We define the mass difference between $2m_\rho$ and $M_1$ as
$\Delta\equiv1-m^2_{\rho}/4M_{1}^2$, and focus on the physical pole $\Delta>0$. 
The cross section should be velocity averaged since the velocity of
DM has a distribution in the DM halo. Assuming the Maxwell-Boltzmann
distribution, the velocity averaged cross section is given by~\cite{Ibe:2008ye} 
\bea
\langle\sigma{v}_{\mathrm{rel}}\rangle=
\frac{1}{2\sqrt{\pi}v_0^3}\int^\infty_0
v^2_{\rm rel} (\sigma v_{\rm rel})e^{-v^2_{\rm rel}/4v_0^2}dv_{\rm rel}.
\eea
where $v_0=10^{-3}$ is the dispersion of the DM velocity. 
The contours of the velocity averaged cross section are shown in Fig.~\ref{mix-plot}. 
The figure suggests, for example, that the required large cross section
$\langle\sigma{v}_{\mathrm{rel}}\rangle\sim10^{-19}~\mathrm{cm^3/s}$ can
be realized when $\Delta$ is $10^{-4}\lesssim\Delta\lesssim10^{-3}$ in the DM mass
range $1~\mathrm{GeV}\lesssim M_1\lesssim100~\mathrm{GeV}$. 
Note that this cross section into $N_2N_2$ does not contribute to estimate
the relic density of DM. 

%%%%%%%%%%%%%%%%%%%
%%%%%%%%%%%%%%%%%%%

After the up-scattering, $N_2$ immediately decays into the ground state DM
($N_1$) and photon through $\eta^{+}$ at one-loop level which is the
dominant decay process of the excited DM. 
The decay width of the process $N_2\to N_1\gamma$ is calculated as
\begin{equation}
\Gamma(N_2\to N_1\gamma)=\frac{\mu_{12}^2}{\pi}\Delta M^3,
\end{equation}
where $\mu_{12}$ is the transition magnetic moment between $N_1$ and
$N_2$ which is calculated as~\cite{Schmidt:2012yg}
\begin{equation}
\mu_{12}\approx\sum_{a}\frac{\mathrm{Im}[(y_\eta^*)_{a1}(y_\eta)_{a2}]eM_1}{2(4\pi)^2m_\eta^2}.
\label{eq:tmm}
\end{equation}
where $e$ is the electromagnetic coupling constant. 
The lifetime of the excited DM $N_2$ should be much less than the
cosmological timescale $\tau\sim10^{17}~\mathrm{s}$ so as to decay immediately
after the $N_2$ production. From the requirement, the order of the Yukawa
coupling $y_\eta$ is estimated as $y_\eta\gtrsim0.01$. 
This does not conflict with the parameter set of Eq.~(\ref{eq:ex-para}). 
As one can see from Eq.~(\ref{eq:tmm}), a complex phase of the Yukawa
coupling $y_\eta$ is necessary to induce the decay $N_2\to N_1\gamma$.
%%%%%%%

\begin{figure}[t]
\begin{center}
\includegraphics[scale=0.6]{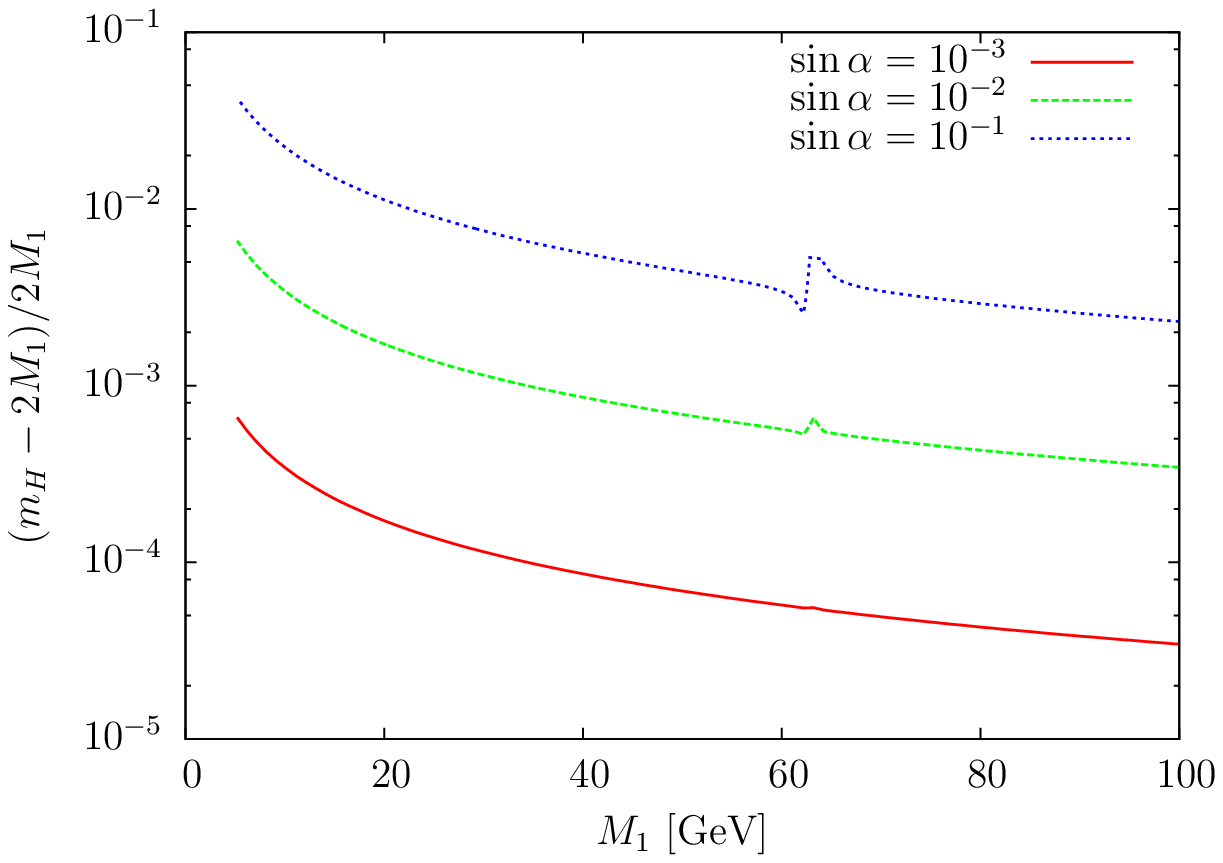}
\includegraphics[scale=0.6]{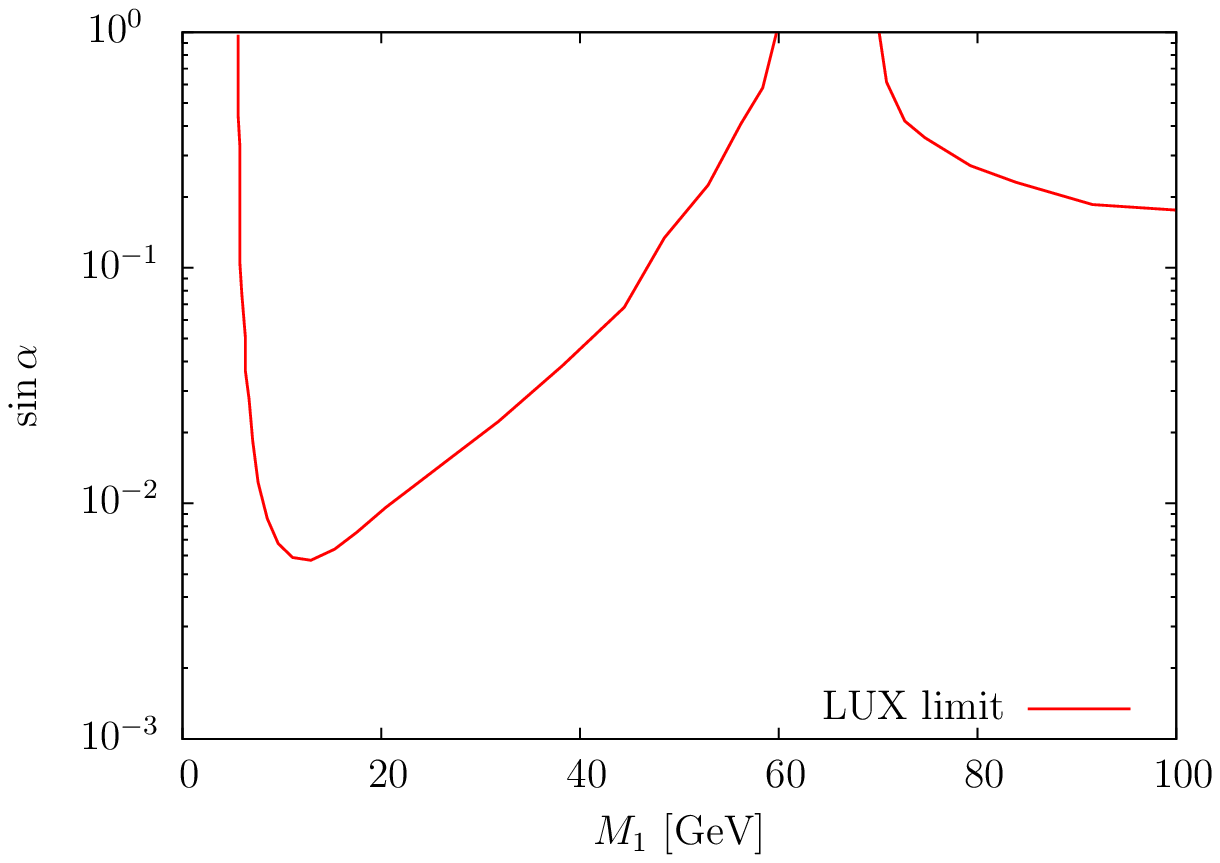}
\caption{Contours of required cross section for thermal relic density of
 DM (left panel) and constraint on mixing angle $\sin\alpha$ from direct
 detection (right panel).}
\label{fig:relic-dd}
\end{center}
\end{figure}

%%%%  Relic Density %%%%
Next we consider the thermal relic density of DM. 
The cross section contributing to the relic density is dominantly given via
$h$ and $H$ s-channel. Although there the other contributions through t and u-channel via
$\eta$ exchange~\cite{Kajiyama:2013rla, Okada:2012np}, these
contributions are negligible due to the heavy mass of the 
intermediate state $\eta$.
Hence we focus on only the s-channel contribution.
The cross section for the channel $N_1N_1\to
f\overline{f}$ mediated by $h$ and $H$ is given by 
\begin{equation}
\sigma v_{\rm rel}
\simeq
\frac{3y_{N1}^2 y_b^2 M_{1}^2\sin^22\alpha}{256\pi} 
\left|\frac{1}{s-m_{h}^2+im_{h}\Gamma_h}
-\frac{1}{s-m_{H}^2+im_{H}\Gamma_H}\right|^2
 \left(1-\frac{m_b^2}{M_{1}^2}\right)^{3/2}v^2_{\rm rel},
\label{eq:relic}
\end{equation}
where only bottom pair is taken into account in
fermion pair $f\overline{f}$ due to the kinematics and strength of
Yukawa coupling. 
As one can see from the equation, we have only p-wave contribution. 
The co-annihilation with $N_2$ should be taken into account since the
masses among them are degenerated. However the order of the effective
cross section including the co-annihilation process is same with
Eq.~(\ref{eq:relic}) as long as $y_{N1}\approx y_{N2}$ is assumed. 
The SM-like Higgs decay width are fixed to be
$\Gamma_h=4.1\times10^{-3}~\mathrm{GeV}$~\cite{higgsdecay}. 
When we consider the mass hierarchy $m_H\gtrsim 2M_1$, 
the dominant decay width of the second Higgs $\Gamma_H$ is expressed as 
\begin{equation}
\Gamma_H=\frac{y_{N1}^2\cos^2\alpha}{16\pi}m_H
\left(1-\frac{4M_{1}^2}{m_H^2}\right)^{3/2}. 
\end{equation}
To obtain the correct relic density $\Omega h^2\approx0.12$, the
required cross section is $\sigma v_{\rm rel}\approx3\times
10^{-26}~\mathrm{cm^3/s}$. 
In the left panel of Fig.~\ref{fig:relic-dd}, the contours of the
required cross section for the thermal relic density are plotted in the
plane of DM mass and the mass degeneracy between DM and $H$. 
As one can see, stronger degeneracy between DM and $H$ is necessary for
smaller mixing angle $\sin\alpha$ in order to induce the appropriate
cross section for the thermal relic density. 
The peak at $M_1\approx63$ GeV is due to the SM Higgs resonance
$2M_1\approx m_h$. 

The direct detection constraint also should be considered since the
scale of our DM is GeV. The elastic cross section with proton is induced
by the t-channel Higgs mediation and it is calculated as 
\begin{equation}
\sigma_p=\frac{Cy_{N1}^2\sin^22\alpha}{8\pi v^2}
\frac{m_p^4M_1^2}{(m_p+M_1)^2}\left(\frac{1}{m_h^2}-\frac{1}{m_H^2}\right)^2,
\end{equation}
where $C\approx0.079$. 
At present, the LUX experiment gives the strongest constraint on the
elastic cross section. 
The constraint of the LUX experiment can be translated to the constraint
on the mixing angle $\sin\alpha$ in our case as shown in the right panel
of Fig.~\ref{fig:relic-dd}. 
In the figure, the mass of the second Higgs $m_H$ is fixed to $m_H=2M_1$
from the requirement of the thermal relic density.
One can see that the mixing angle $\sin\alpha$ should be $\sin\alpha\lesssim0.005$ in 
order to evade the direct detection constraint in whole DM mass range. 

%Its value can be achieved when we provide the following parameter set for example:
%\be
%y_{N1}\approx1,\quad \alpha\approx10^{-4},\quad m_H\approx120\ {\rm GeV},
%\ee 
%where the others are experimentally fixed. 
%In this parameter set, the constraint of the direct detection can be
%avoided because its spin independent elastic scattering cross section is
%around $\sigma_{\mathrm{SI}}\sim10^{-48}$ cm$^2$ at $M_{N1}\approx10$ GeV,
%which is lower than the current experimental bound such as LUX whose
%upper bound is $\sigma_{\mathrm{SI}}\lesssim10^{-44}$ cm$^2$. 
%Notice here that the constraint from  the invisible decay of the SM
%Higgs is always milder than the one from the direct detection. Hence
%this constraint is always safe in our model as long as the constraint
%of the direct detection is satisfied.  

\section{Summary and Conclusion}
We have studied an exciting DM scenario in a radiative neutrino model to explain
the X-ray line signal at $3.55$ keV recently reported by XMN-Newton
X-ray observatory using data of various galaxy clusters and Andromeda galaxy. 
We have shown that neutrino masses can be radiatively generated
by our DM with the mass of ${\cal O}$(10) GeV, which is requested by the
exciting DM scenario. 
Also we have shown that the required large cross section to explain the
X-ray line can be obtained by the resonance of the massive pseudo scalar
$\rho$ that provides s-wave contribution for only the up-scattering
process $N_1N_1\to N_2N_2$. 
The model can be consistent with the observed relic density as well as the
direct detection constraint. 
To induce 3.55 keV X-ray line without any inconsistencies, we have found
that the mass degeneracies $m_H\approx m_\rho\approx2M_1$ are required in the model.

%\fi

%\newpage
%%%%%%%%%%%%%%%%%%%%%%%%%%%%%%%%%%%
%\vspace{0.5cm}
%\hspace{0.2cm} {\bf Acknowledgments}
\section*{Acknowledgments}
\vspace{0.5cm}
H.O. thank to Prof. Seungwon Baek for fruitful discussions. 
T.T. acknowledges support from the European ITN project (FP7-
PEOPLE-2011-ITN, PITN-GA-2011-289442-INVISIBLES). 
%%%%%%%%%%%%%%%%%%%%%%%%%%%%%%%%%%%

\end{document}